\documentclass{iaus}
\usepackage{graphicx}

\title[Spectroscopic Monitoring of LBVs] 
{Long-term Spectroscopic Monitoring of LBVs and LBV Candidates}

\author[Lobel et al.]   
{A. Lobel$^1$, J. H. Groh$^2$, K. Torres$^1$ \and N. Gorlova$^3$}

\affiliation{$^1$Royal Observatory of Belgium \\ Ringlaan 3,
B-1180, Brussels, Belgium \\ email: {\tt Alex.Lobel@oma.be; alobel@sdf.lonestar.org; Kelly.Torres@oma.be} 
\\[\affilskip]
$^2$Max Planck Institute for Radio Astronomy, \\ Auf dem H\"{u}gel 69, 53121 Bonn, 
Germany \\email: {\tt jgroh@mpifr-bonn.mpg.de} \\[\affilskip]
$^3$Institute of Astronomy, Katholieke Universiteit Leuven, \\ 
Celestijnenlaan 200D BUS 2401, 3001 Leuven, Belgium \\ 
email: {\tt nadya@ster.kuleuven.be}
}

\pubyear{2010}
\volume{xxx}  
\pagerange{1--2}
\jname{Title of your IAU Symposium}
\editors{A.C. Editor, B.D. Editor \& C.E. Editor, eds.}
\begin{document}

\maketitle

\begin{abstract}
We present results of a long-term spectroscopic monitoring program (since mid 2009) of 
Luminous Blue Variables with the new HERMES echelle spectrograph on the 1.2m Mercator telescope 
at La Palma (Spain). We investigate high-resolution ($R$=80,000) optical spectra of two LBVs, 
P~Cyg and HD~168607, the LBV candidates MWC~930 and HD~168625, and the LBV binary MWC~314. 
In P Cyg we observe flux changes in the violet wings of the Balmer H$\alpha$, H$\beta$, 
and He~{\sc i} lines between May and Sep 2009. The changes around 200 to 300 
$\rm km\,s^{-1}$ are caused by variable opacity at the base of the supersonic wind from 
the blue supergiant. \\ 
We observe in MWC~314 broad double-peaked metal emission lines with invariable radial 
velocities over time. On the other hand, we measure in the photospheric S~{\sc ii} $\lambda$5647 
absorption line, with lower excitation energy of $\sim$14~eV, an increase of the heliocentric radial 
velocity centroid from 37~$\rm km\,s^{-1}$ to 70~$\rm km\,s^{-1}$ between 5 and 10 Sep 2009 
(and 43~$\rm km\,s^{-1}$ on 6 Apr 2010). The increase of radial velocity of $\sim$33~$\rm km\,s^{-1}$ 
in only 5 days can confirm the binary nature of this LBV close to the Eddington luminosity limit. \\
A comparison with VLT-UVES and Keck-Hires spectra observed over the past 13 years reveals strong 
flux variability in the violet wing of the H$\alpha$ emission line of HD~168625 and in the absorption 
portion of the H$\beta$ line of HD~168607. In HD~168625 we observe H$\alpha$ wind absorption at 
velocities exceeding 200~$\rm km\,s^{-1}$ which develops between Apr and June 2010.
\end{abstract}

\firstsection 
\section{Introduction}
The HERMES instrument on the 1.2m Mercator telescope at La Palma 
(\cite[Raskin \& Van Winckel 2008]{Raski08}) is a new high-efficiency fiber-fed 
bench-mounted cross-dispersed echelle spectrograph that observes the complete wavelength range from 
420 nm to 900 nm in a single exposure with $R$=80,000. TAC reviewed HERMES observation programs of 
the contributing research institutions started mid 2009. We present first results  
of a long-term high-resolution spectroscopic monitoring program of 3 LBVs and 2 LBV candidates (up 
to $V$=$11^{\rm m}.0$). The HERMES monitoring program will provide invaluable new 
clues about the structure and dynamics of LBV atmospheres, the physics of their extended winds, and 
the strong line broadening mechanisms in these rare massive hot stars near the Eddington luminosity 
limit. The monitoring program will be crucial for documenting the enigmatic LBV outburst events, for 
detecting new long-period LBV binaries and the reliable determination of the orbital parameters.

\section{The LBV Binary MWC~314 and LBV Candidate MWC~930}

\begin{figure}[t]
\begin{center}
 \includegraphics[width=5.4in]{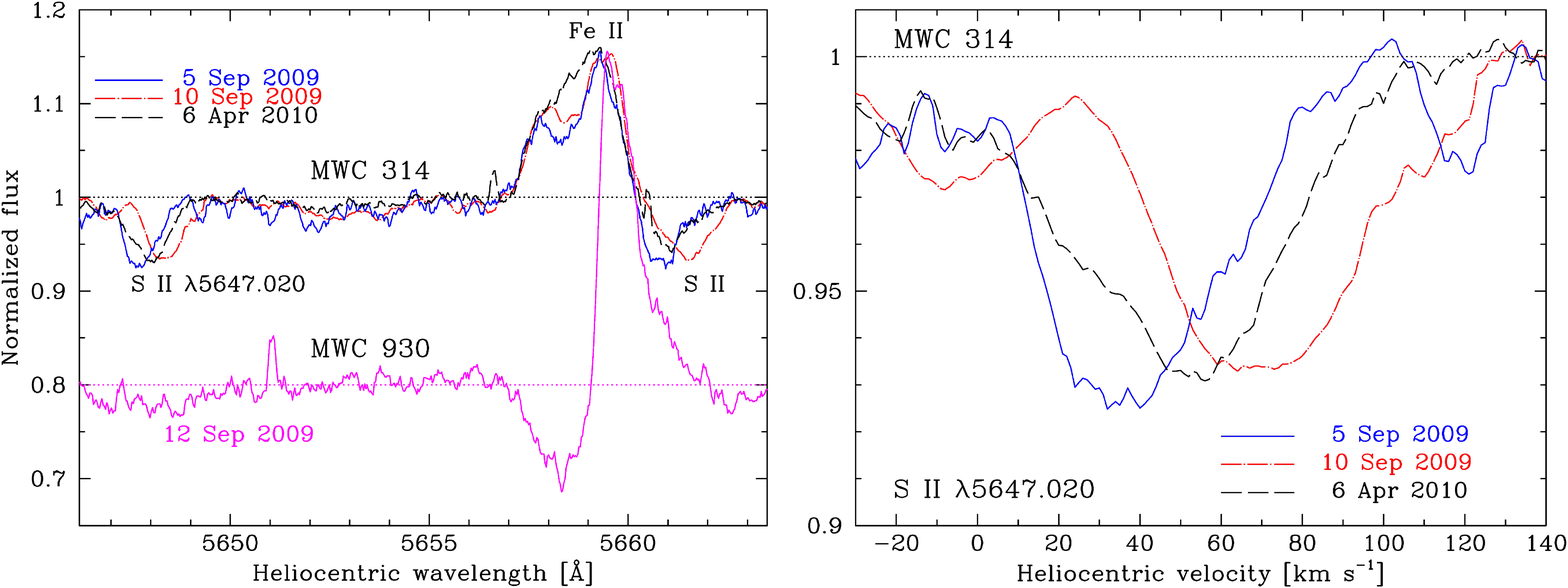} 
\caption{Large radial velocity changes of $\sim$33~$\rm km\,s^{-1}$ 
observed in S~{\sc ii} lines of MWC~314 over 5 days with Mercator-HERMES 
confirm its LBV binary nature ({\it see text}).   }
\label{fig1}
\end{center}
\end{figure}

MWC~314 was observed with HERMES on 5 and 10 Sep 2009, and on 6 April 2010. 
\cite[Muratorio et al. (2008)]{Murat08} found evidence for a $\sim$30 day orbital 
period and proposed that its double-peaked permitted and forbidden ionic emission 
lines (Fe~{\sc ii}, Cr~{\sc ii}, Ti~{\sc ii}) are formed in a warm disk rotating 
around the star. Its orbital parameters are however still unknown at present.
We observe with HERMES many broad and double-peaked optical emission lines 
(i.e., Fe~{\sc ii} $\lambda$5657) reminiscent of a Be-star spectrum. 
The emission line flux maxima are strongly variable due to wind opacity changes 
in the emission line formation region. 
The radial velocities of these optical emission lines are however invariable, 
signaling a stable circumstellar or circumbinary (disc) envelope. We observe 
large changes in the radial velocity centroid of photospheric S~{\sc ii} absorption 
lines from 37~$\rm km\,s^{-1}$ ({\it blue curve in the right-hand panel of 
Fig.\,\ref{fig1}}) to 70~$\rm km\,s^{-1}$ ({\it red curve}) in only 5 days. 
Our short-term spectroscopic observations with HERMES hence confirm the binarity 
of MWC~314. We observe many strong optical P Cygni profiles in the LBV 
candidate MWC~930 in Sep 2009 ({\it magenta curve in Fig.\,\ref{fig1}}) 
signaling the wind properties of a massive hot star. 

\section{Summary}
Long-term monitoring with Mercator-HERMES of the optical spectrum of the prototypical LBV
P Cyg reveals variability at the base of its supersonic wind that can be linked to moderate 
$V$-changes (of $0^{\rm m}.1$ to $0^{\rm m}.2$) over a period of $\sim$4 months. 
We find strong indications for the binary nature of MWC~314 from large radial 
velocity changes observed in photospheric absorption 
lines during less than one week. We observe prominent P Cygni profiles in the optical 
spectrum of LBV candidate MWC~930, signaling the presence of a central massive hot star. 
The optical spectral lines of LBV candidate HD~168625 are less variable, although we 
also observe clear signatures of expanding H$\alpha$ wind variability on 
short time-scales of $\sim$1 month. The optical spectrum of HD~168607 reveals large line 
profile changes over the past 12 years confirming its LBV designation.


\begin{thebibliography}{}

\bibitem[Raskin, G. \& Van Winckel, H. (2008)]{Raski08}  
{Raskin, G., \& Van Winckel, H.} 2008, 
\textit{SPIE}, 7014, 178 

\bibitem[Muratorio, R., Rossi, C., \& Friedjung, M. (2008)]{Murat08}
{Muratorio, R, Rossi, C., \& Friedjung, M.} 2008,
\textit{A\&A}, 487, 637

\end{thebibliography}
\end{document}